\documentclass[aps,prl,twocolumn,superscriptaddress,showpacs]{revtex4-1}

\usepackage{graphicx}
\usepackage{color}
\usepackage{amsbsy,amssymb,amsmath,bm}
\usepackage{float}
\usepackage{verbatim}

\begin{document}

\title{Intrinsic non-linear conduction in the super-insulating state of thin TiN films}

\author{D. Kalok}
\affiliation{Institute of Experimental and Applied Physics, University of Regensburg, D-93025 Regensburg, Germany}
\author{A.\,Bilu\v{s}i\'{c}}
\affiliation{Institute of Experimental and Applied Physics, University of Regensburg, D-93025 Regensburg, Germany}
\affiliation{Department of Physics, Faculty of Science, University of Split, HR-21000 Split, Croatia}
\author{T.\,I. Baturina}
\affiliation{Institute of Semiconductor Physics, 13 Lavrentjev Avenue, Novosibirsk, 630090 Russia}
\affiliation{Materials Science Division, Argonne National Laboratory, Argonne, Illinois 60439, USA}
\author{V.\,M. Vinokur}
\affiliation{Materials Science Division, Argonne National Laboratory, Argonne, Illinois 60439, USA}
\author{C. Strunk}
\affiliation{Institute of Experimental and Applied Physics, University of Regensburg, D-93025 Regensburg, Germany}
\email[]{christoph-.strunk@physik.uni-regensburg.de}

\date{\today}

\begin{abstract}
We investigate experimentally the electric transport at the insulating side of the superconductor to insulator
transition in thin TiN-films. At temperatures $T > 50$\,mK we observe an Arrhenius-type conductance,
with an activation energy depending logarithmically on the sample size. At high bias the current voltage ($I$-$V$) characteristics display a large current jump into an electron heating dominated regime.
 For the largest samples, and below 50\,mK  we observe a low-bias power law $I\propto V^\alpha$ characteristics with an exponent $\alpha>1$ rapidly growing with decreasing temperature, which is expected for a binding-unbinding crossover of the charge-Berezinskii-Kosterlitz-Thouless type.
\end{abstract}

\pacs{74.25.fc, 73.50.-h, 74.25.Dw, 74.78.-w}


\maketitle

An intriguing low-temperature highly
resistive state forms in disordered InO and TiN films
approaching  the disorder-driven
superconductor-insulator transition (D-SIT) from its insulating side~\cite{Shahar2005,TiNIns}.
A characteristic feature of this state is the switching character of the current-voltage ($I$-$V$)
characteristics: at certain magnetic field ($B$)-dependent threshold voltage $V_{HL}$
the current jumps up over several orders of magnitude.
It was further found that this highly
resistive state in TiN can show hyperactivated temperature behavior of the resistance~\cite{TiN_HA}.

Comparing the data on the TiN films and the two-dimensional Josephson junction arrays (JJA)~\cite{Kanda}, and exploiting
the duality between the superconducting and insulating states in JJA~\cite{Fazio, Mooij1990}, it was proposed~\cite{FVB,VinNature}
that in disordered TiN films a distinct superinsulating state forms, which is a low temperature phase of the charge
Berezinskii-Kosterlitz-Thouless (BKT) transition. Within the JJA-model conduction in the insulating state is caused by tunneling of Cooper pairs from one island to the next, leaving an opposite charge behind. At low temperatures long range Coulomb interactions start to dominate the conduction process: a single charge/anti-charge pair generates polarization charges all over the sample, which requires an energy that increases in two dimensions logarithmically with the size of the sample
(provided the electrostatic screening length exceeds the sample size). Hence the thermal activation energy becomes dependent on the sample size~\cite{FVB,VinNature}.
The tunneling requires a
cascade energy relaxation via the emission of electric dipole excitations and phonons. The microscopic mechanism of superinsulation was proposed to be the suppression of this relaxation and thus blockade of the tunneling current below the charge BKT transition~\cite{CVB}.

 A detailed study of the current jumps in InO was undertaken in~\cite{Ovadia}.  It was found that the $I$-$V$
isotherms measured on the magnetic field-tuned insulator at $B=11$\,T  above the $B$ value of the maximum of the
magnetoresistance peak that, for this sample, is at $B=8$\,T, and slightly exceeding the upper critical field, $B_{c2}$,
are described excellently by the standard heating instability model~\cite{Volkov_Kogan,Clarke1989,Giazotto}.  The latter was
adapted~\cite{Altshuler} to strongly disordered materials by substituting into general expressions of ~\cite{Volkov_Kogan,Clarke1989,Giazotto}
the electron-phonon relaxation rates calculated for strongly
disordered conductors in~\cite{Schmid,Reizer-JETP,Reizer1989,Sergeev}
(see~\cite{Giazotto} for a comprehensive review).
%
%

The findings of~\cite{Ovadia} presented convincing evidence that at high fields
Cooper pairing the InO films is destroyed and these films behave like
conventional strong insulators. On the other hand,  fundamental questions calling for detailed experimental investigation remain:
(i) The nature of the high-resistive state near the D-SIT where a Cooper pair insulator (CPI) forms,
(ii) The role, if any, of the long range Coulomb interaction in formation CPI properties, and
(iii) The role of charge BKT in formation of the high resistive state of CPI.

In this Letter, we address these questions.  We measure the activation energy in the CPI-state, and find it indeed depending logarithmically upon the sample size.
Our data reveal a charge BKT-like cross-over, which marks the formation of a low temperature state with highly nonlinear $I$-$V$ curves. Below $50\,$mK $I(V)$ follows a power-law $I\propto V^\alpha$ behavior with an exponent $\alpha$ rapidly growing with the decrease of temperature, which
provides evidence for super-insulating behavior at very low temperatures. Finally, we uncover a broad distribution of switching voltages from the HR- to the low-resistive (LR) state at low temperatures
in a strong contrast to switching governed by the overheating instability observed in~\cite{Ovadia}.

We measured the $I$-$V$-characteristics on the
5\,nm thin disordered TiN films on the verge of the D-SIT similar by their parameters, to those studied in~\cite{TiNIns,TiN_HA,TiN_QM}.
The data presented below are taken on two samples: the sample S which is superconducting at zero magnetic field
but can be driven into an insulating state by applying a small magnetic field, and the sample I, remaining insulating even at $B=0$.
Magnetic field was taken perpendicular to the sample plane.
The sample S has a Hall-bar form. All $I$-$V$s were measured on a stripe connecting the two most distant contacts
(13 squares corresponding to 650 $\mu$m and 50 $\mu$m in length and width, respectively).
The transition temperature at zero magnetic field is below 100~mK.
On a different chip of the same wafer, the sample I was patterned into the six independent squares of different side lengths,
$L$ ($L=0.5, 2, 5, 20, 240$, and $500$\,$\mu$m), and attached to two wide and thick gold pads, to provide a uniform current distribution
and low contact resistances (see inset in Fig.~1b). After the patterning process these samples were
found insulating for $T \rightarrow 0$\,K at $B=0$\,T.
The results presented below are based on the detailed analysis of about 4000 dc $I$-$V$
characteristics measured by the two-probe technique.
Bias voltages were provided by a dc voltage source Yokogawa 7651 via a voltage divider (1000:1), with a highest sweeping rate of 1\,mV/min.
Currents were measured with DL1211 and Femto DDPCA-S current preamplifiers.
In the superconducting regime current biased four-probe measurements were carried out; in the insulating regime two-probe voltage biased measurements were sufficient as the contact resistance was negligible below 1~K.
The used setup enabled us to measure resistances exceeding 100\,G$\Omega$.

\begin{figure}[t]
\includegraphics[width=86mm]{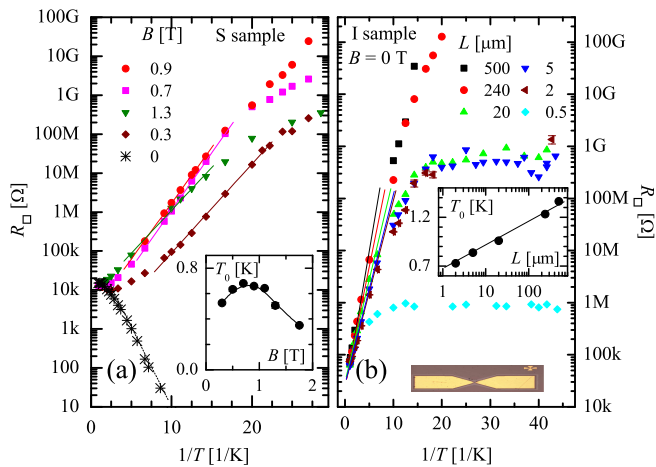}
\caption{\label{fig: RTBi} (color online)
(a) $R_{\square}$ vs. $1/T$ for Sample S at different magnetic fields $B$. Inset: Activation energy $T_0$ vs.~$B$  of sample S. Line is guide for an eye. (b) Sample I for different sample lengths at zero magnetic field. Insets: $T_0$ vs.~sample size $L$ of Sample I at zero magnetic field (top) and optical image of the $5 \times 5\,\mu$m device of sample I (bottom). Solid lines on both main panels are fits to Arrhenius law.
}
\end{figure}

We first discuss the effects of a magnetic field on the sheet resistance of the S-sample.  In the
Fig.~\ref{fig: RTBi}a we show $\log R_{\square}$ vs. $1/T$ plots where each data point depicts the zero-bias resistance
extracted from the dc $I$-$V$-characteristics in the linear response regime.
The superconducting state of the Sample S is very fragile and suppressed by small magnetic fields $B\lesssim 5\;$mT. Solid lines represent fits to an Arrhenius law ($R_{\square}(T) = R_{0} \exp{(T_0/T)}$, where $k_B T_0$ is an activation energy).
 At the lowest temperatures $R(T)$ deviates from the Arrhenius behavior and for $T\lesssim50\;$mK no linear resistance could be extracted from the $I$-$V$-curves (see below). In parallel magnetic field qualitatively similar behavior  was found with two orders of magnitude smaller typical resistances (not shown). The magnetic field dependence of $T_0$  exhibits a nonmonotonic $T$-dependence with a maximum around 0.9~T similar to earlier experiments \cite{TiNIns,TiN_HA,TiN_QM}, and is plotted in the inset of Fig.~\ref{fig: RTBi}a. Employing Josephson-junctions network (JJN) model~\cite{FVB} to $T_0 (B)$, one can estimate $d_\textrm{S} = 28$\,nm for the size of the Josephson junctions plaquettes for Sample S .

In Fig.~\ref{fig: RTBi}b we present $\log R_\square (1/T)$ of squares with increasing side length $L$ for Sample I.
The $R_\square (1/T)$ curves display again Arrhenius behavior at higher temperatures, but at low $T$ striking differences occur: while the largest sample appear to follow the Arrhenius law down to the lowest temperatures, for which a linear resistance can be extracted, the smaller samples exhibit a saturation of $R_\square$ at low $T$, the saturation level strongly depending on the sample size. However, also in the high-$T$ region differences exist:
the inset to Fig.\,\ref{fig: RTBi} (b) presents $L$ dependence of $T_0$ for Sample I at zero magnetic field. The observed linearity of $T_0$ vs.~$\ln L$ is consistent with the JJN model, and invalidates the concept of resistivity as a geometry independent characteristic of the TiN film. It resembles that of the InO$_x$ films in \cite{kowal_ovadyahu} and strongly supports the idea of a collective Coulomb energy introduced in \cite{FVB}. Using Eq.~12 of \cite{FVB} the data provide a single island charging energy $E_{c\textrm{I}}/k_B=0.2\,$K.

\begin{figure}[t]
\includegraphics[width=86mm]{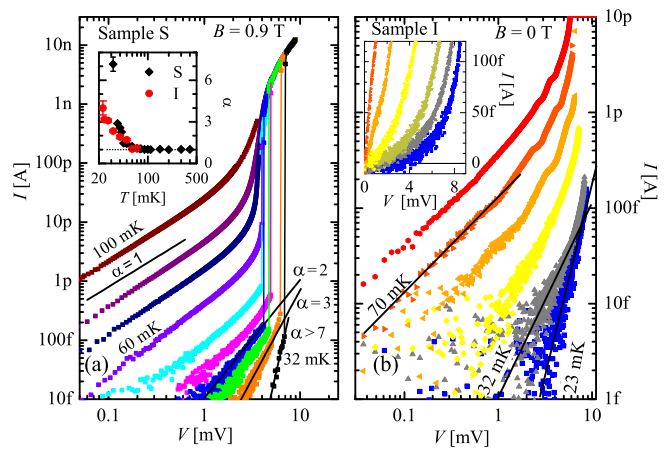}
\caption{\label{fig: logI-logV}  (color online)
$I$-$V$ characteristics (a) of sample S at $B = 0.9$\,T and
$T=100,80,70,60,50,45,42,40,37$ and $32\,$mK from top to bottom; and
(b) of sample I at $B=0$\,T and $T=80,70,60,50,40,32$ and $24$\,mK and $L=240\,\mu$m.
The voltage is swept from negative to positive values,
and only a positive voltages are presented.  Solid lines indicate
the slopes corresponding to different values of power $\alpha$ in the
$I\propto V^\alpha$ dependence. Insets in (a) and (b) display $\alpha(T)$ for both samples and $I$-$V$ in linear scale for Sample I, respectively.
}
\end{figure}

Our central result is presented in Fig.\,\ref{fig: logI-logV} where we plot representative $\log I$ vs. $\log V$ traces
for sample S ($B = 0.9$\,T), and Sample I $(B=0\;$T).
In either sample $I$-$V$s display switching from the highly resistive (HR)-
to low resistive (LR) state at the threshold voltage $V_{\mathrm{ HL}}$
(we follow convenient notations introduced in~\cite{Ovadia}).
The current jumps are seen over rather wide $B$-range (from
0.3 to 1.5 T for Sample S).
Note that these fields are well below the upper critical field
in TiN superconducting films, $B_{\mathrm{c2}}(0)=2.8$\,T~\cite{TiN_SMT}.
At higher temperatures $T\gtrsim 70\,$mK in both samples $I$-$V$s are linear (ohmic) at low voltages but
develop a characteristic steep, but continuous increase with increasing voltage.
At high voltages all $I$-$V$s for different temperatures merge.
 Below $T\approx 50\;$mK the switching voltages increase on average, but begin to scatter randomly over a finite temperature dependent $V$-interval (see below).

Most striking is that both samples exhibit a power-law $I\propto V^\alpha$ behavior with a dramatically
growing exponent $\alpha(T)$  in the pre-switching regime (see the inset in Fig.~\ref{fig: logI-logV}a). E.g.~for sample S
$\alpha=2$ at $T=42$\,mK, reaching
the value $\alpha(T_{\scriptscriptstyle{\mathrm{BKT}}})=3$ at $T=37$\,mK
and shooting up beyond seven at $T\lesssim 32$\,mK.
At $T<32$\,mK the HR parts of $I$-$V$s are concealed under the noise floor
and a finite voltage springs up
at $V_{\mathrm{HL}}$ in the LR part of the $I$-$V$ curves.
Such an evolution of the $I$-$V$ curves with decreasing temperature and the corresponding $T$-dependence of $\alpha$  is characteristic of a charge BKT-transition which is expected to occur for
Josephson networks in the insulating regime, as described in the introduction~\cite{Mooij1990,Fazio}, since
the number of free charges depends on voltage, leading to an intrinsically non-linear $I$-$V$ characteristics.
Thus our present findings lend strong support to the idea~\cite{FVB,VinNature} that
at very low $T$ a crossover to a  superinsulating state exists, which is of the charge BKT nature.

\begin{figure}[t]
\includegraphics[width=85mm]{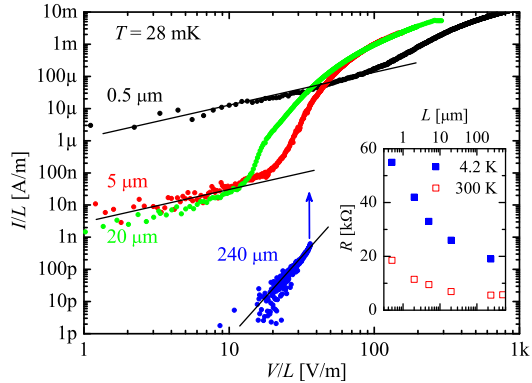}
\caption{\label{fig: size_dep}  (color online) Normalized $I$-$V$-curves at 28 ~mK and different sample sizes. The lines indicate power law behavior for $\alpha=1$ for $L=0.5, 5,$ and 20$\,\mu$m, and $\alpha\simeq5$ for 240$\,\mu$m.
Inset: $R(L)$ at higher temperatures.}
\end{figure}
We now address the size dependence of the $I$-$V$ characteristics at low temperature shown in Fig.~\ref{fig: size_dep}. For better comparison the $I$ and $V$ axis are normalized by the the sample size $L$. In absence of long range Coulomb effects the
sheet resistance should be size independent and all curves in Fig.~\ref{fig: size_dep} should collapse. This is clearly not the case: the $I$-$V$ characteristics for the three the shortest samples is linear at low $V$, but with a very different resistivity. The size dependence of the $I$-$V$ curves and the saturation of the sheet resistance (cf. Fig.~\ref{fig: logI-logV}b) at the lowest temperature depends is much more dramatic than expected from the weak logarithmic size dependence of the activation energy.
At higher $V$ the a step-like feature occurs, which is most probably heating induced. The corresponding electric field increases with decreasing sample size. Only the largest sample displays a strongly non-linear $I$-$V$ characteristic at low $V$ with a very high power-law exponent $\alpha\simeq5$. The inset displays $R(L)$ at higher temperatures $T>4\;$K, where the size dependence is reverted: $R$ decreases with $L$. While the precise reason for this decrease is not yet clear, it is plausible that the patterning process, which has already driven the sample towards the insulating side of the SIT, affects in particular the sample edges. A locally suppressed conductivity at the sample edges affects more strongly the smaller samples. An overall gradient of the local conductivity over the whole chip can be excluded, since the smallest and the largest sample are placed next to each other.

We now discuss the properties of the sharp current jumps between the HR and the LR state. The latter had been attributed to a dielectric breakdown along a
quasi-1D weakest dielectric strength channel~\cite{FVB}. While a random distribution of $V_{HL}$ is typical for dielectric breakdown phenomena \cite{dielectrics}, a complete description of threshold behavior requires in addition the consideration of heating instabilities~\cite{Volkov_Kogan,ReviewHeating} mentioned in the introduction, which usually accompany a dielectric breakdown.
Such an electron overheating governs in particular the return from the LR to the HR state upon reducing the bias voltage. Adopting the model considerations in Ref.~\cite{Ovadia} we find that we can fit the $I$-$V$ characteristics
of Fig.~\ref{fig: logI-logV} for $T\gtrsim 60\,$mK \cite{CVB_heating}.
In contrast, at $T<60\,$mK the heating model significant overestimates the range of measured $V_{HL}$, indicating that the heating instability is preempted by a dielectric breakdown at the lowest $T$.

\begin{figure}[t]
\includegraphics[width=85mm]{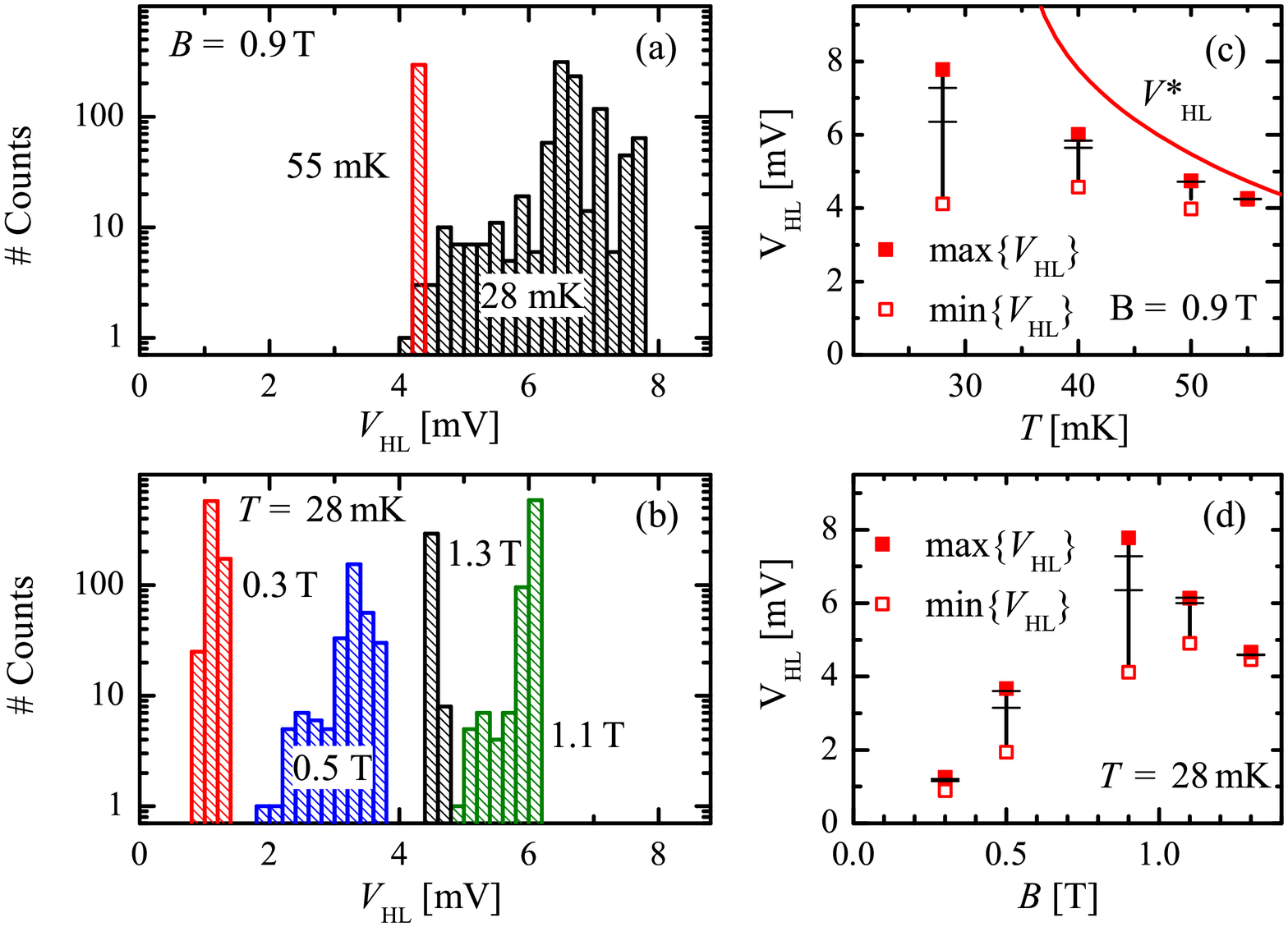}
\caption{\label{fig: jittering}  (color online)
Histograms for the $V_{\mathrm{HL}}$ distribution:
(a)  $ = 0.9$\,T
and $T = 28$\,mK (the total sampling number is 930)
and $T = 55$\,mK (296 samplings);
(b) $T = 28$\,mK and  $B = 0.3$\,T (776 samplings),
0.5\,T (299 samplings), 1.1\,T (708 samplings), and 1.3\,T (300 samplings).
(c)-(d): Temperature (at $ = 0.9$\,T) and field (at $T = 28$\,mK)
dependences of the variance range of $V_{\mathrm{HL}}$,
maximal and minimal values shown by solid and open symbols, respectively.
Horizontal lines indicate intervals containing 75\% of the data.
The solid line marked by $V_{\mathrm{HL}}^{\ast}$ corresponds
to the conventional overheating model.
}\end{figure}

To investigate the statistical distribution of the switching voltages $V_{\mathrm{HL}}$
at $T<60$\,mK, we have measured histograms of the $I$-$V$ characteristics for each representative temperature and external magnetic field.
Figure~\ref{fig: jittering} displays the corresponding histograms of $V_{\mathrm{HL}}$ for Sample S
[panels (a) and (b)];
panels (c) and (d) show the temperature and the magnetic field dependence
of the maximal and minimal values of $V_{\textrm{HL}}$.  We observe that i) the histograms display a sharp cut off at the maximal observed $V_{\mathrm{HL}}$ and is asymmetric with respect to the most probable $V_{\mathrm{HL}}(B)$ (see the representative histogram for $B = 0.9$\,T and $T=28$\,mK). ii) The field-dependence of the most probable value of $V_{\mathrm{HL}}(B)$ is non-monotonic and has the values close to those observed on samples which are insulating at $B=0$~\cite{TiNIns}. iii) The ratio of max\{$V_{\mathrm{HL}}(B)$\}/min\{$V_{\mathrm{HL}}(B)\} \approx 2$
in magnetic fields below the field where the maximum of $V_{\mathrm{HL}}(B)$ is achieved,
and drops rapidly upon further field increase.
iv) As temperature grows, max\{$V_{\mathrm{HL}}$\}/min\{$V_{\mathrm{HL}}$\} shrinks to unity and the random scattering of $V_{\mathrm{HL}}$ vanishes. Note that the variance of
$V_{\mathrm{HL}}$ disappear much faster than $V_{\mathrm{HL}}$
itself. The observed jitter in the threshold voltage closely resembles the dielectric breakdown behavior
in conventional semiconductors.
A similar random switching $I$-$V$ behavior with the characteristic current jumps
and hysteresis has been also observed in
 2D Josephson junction networks
in the insulating state~\cite{JJArray1,JJArray2,Haviland}.

In conclusion the above observations reveal two
characteristically different regimes within the insulating side of the SIT in our TiN films:
i) at comparatively high temperatures 60~mK~$\lesssim T\lesssim$~500~mK transport is linear with a thermally activated resistance.
The activation energy increases with the sample size. The evolution with magnetic field and size can be consistently explained within a Josephson network model in the insulating regime.
 As local Cooper pairing is essential for the insulating behavior, we call this regime a {\it Cooper pair insulator}.
ii) at very low temperature $T\lesssim50\;$mK the conduction process becomes intrinsically non-linear at low bias voltage.

Hence, our low-temperature data are in accord with the existence of the low temperature charge Berezhinskii-Kosterlitz-Thouless phase, supporting the notion of {\it superinsulation} as dual to superconductivity. Measurements of the dielectric properties could provide independent and complementary evidence for this new state of matter.

\begin{acknowledgments} We thank M.~Baklanov for the supply of TiN material.
The work was supported by the Deutsche Forschungsgemeinschaft under Grant No. 444USA113/3 and
within the GRK 638,
the Russian Foundation for Basic Research (Grant No. 09-02-01205),
and the U.S. Department of Energy Office of Science under the Contract No. DE-AC02-06CH11357.
\end{acknowledgments}

\end{document}